\documentclass{article}

\newcommand{\mq}{\mathbf{q}}
\newcommand{\mf}{\mathbf{f}}
\newcommand{\mpsi}{\boldsymbol{\psi}}

\newcommand{\mA}{\boldsymbol{A}}
\newcommand{\mJ}{\boldsymbol{J}}

\newcommand{\p}[1]{(\ref{#1})}

\newcommand{\cF}{{\cal F}}

\newcommand{\cD}{{\cal D}}

\newcommand{\cA}{{\cal A}}

\newcommand{\cE}{{\cal E}}
\newcommand{\cL}{{\cal L}}

\newcommand{\bpsi}{{\bar\psi}{}}

\newcommand{\bq}{{\bar q}}

\newcommand{\be}{\begin{equation}}
\newcommand{\ee}{\end{equation}}
\newcommand{\bea}{\begin{eqnarray}}
\newcommand{\eea}{\end{eqnarray}}

\newcommand{\ba}{\begin{array}} \newcommand{\ea}{\end{array}}

\def\im{{\rm i}}
\newcommand{\tr}{{\rm Tr}{\,}}

\newcommand{\nn}{\nonumber}

\usepackage{amscd,amsmath,amssymb}

\begin{document}
\thispagestyle{empty}
\vspace{2cm}
\begin{flushright}
\end{flushright}\vspace{2cm}
\begin{center}
{\Large\bf Partial breaking of arbitrary amount\\[1cm]
of $d=3$ supersymmetry}
\end{center}
\vspace{1cm}

\begin{center}
{\large\bf  N.~Kozyrev${}^a$}
\end{center}

\begin{center}
${}^a$ {\it
Bogoliubov  Laboratory of Theoretical Physics, JINR,
141980 Dubna, Russia} \vspace{0.2cm}
\end{center}
\vspace{2cm}

\begin{abstract}\noindent
Among the solutions of string theory and supergravity which preserve some fraction of supersymmetry, the best known are those that leave one half of the supersymmetry unbroken, and there is a large number of field theory models with this pattern of supersymmetry breaking. However, a lot of brane configurations exist which preserve only $1/4$, $1/8$ or more exotic fractions of supersymmetry, and field theory side of these systems remains largely unexplored. To find whether the formalism of nonlinear realizations is useful in construction of models of this type, we consider the systems of some $N_0$ scalar and vector $N=1$, $d=3$ Goldstone supermultiplets. We find that it is possible to construct an $SO(N_0)$ invariant theory of $N_0$ scalar multiplets with $N_0$ broken supersymmetries. For $N_0=3$ or $N_0\geq 5$ its action is not of Nambu-Goto type and its structure remains universal for arbitrary $N_0$. The cases of $N_0=1,2$ correspond to the membranes in $D=4$ and $D=5$, respectively, while for $N_0=4$ some arbitrariness in the action remains, and with proper choice of parameters, it is possible to obtain the action of the membrane in $D=7$ in the bosonic limit. It is also shown that the $SO(N_0)$ invariant action of $N_0$ vector multiplets with $1/N_0$ pattern of supersymmetry breaking does not exist for arbitrary $N_0$.

\end{abstract}

\setcounter{page}{1}
\setcounter{equation}{0}


\section{Introduction}
Studies of systems with partial spontaneous breaking of global supersymmetry make up an important branch of all supersymmetry studies. Systems with PBGS play an important role in string theory and supergravity where a lot of solutions that satisfy the BPS condition and preserve some fraction of supersymmetry can be found. The best known solutions preserve one half of the supersymmetry and comprise the brane scan \cite{branescan,12pbgs,stellerew}. For many of these solutions (D-branes, p-branes) the effective actions, superfield or component, which can be determined by the invariance with respect to both broken and unbroken supersymmetries, are known. However, there are a lot of other solutions with different patterns of supersymmetry breaking, which correspond to intersecting branes. Solutions were found that preserve the $n/32$ fraction of supersymmetry, where $n=0,1,2,3,4,5,6,8,16$ (see \cite{34pbgs}, where also the $3/4$ fraction was proposed, and references therein). For systems with these patterns of breaking, the effective actions are much less known. These include particles with tensorial central charges \cite{14pbgs1,14pbgs2,FedZima},  but attempts to construct field theory actions were only briefly outlined and were not truly successful \cite{d10break}. The $\kappa$-symmetric theories with the $3/4$ and even the $\frac{n-1}{n}$ patterns of breaking can be found in \cite{34pbgsZU,BengZhel}. Thus, it would be of interest to try to employ the formalism of nonlinear realizations to construct component actions for systems with unusual patterns of supersymmetry breaking extending the ideas of works \cite{cosetact}, \cite{actD5}.

The idea of constructing component actions of theories with partial breaking of supersymmetry is related to the fact that one can realize transformations of the superspace coordinates, Goldstone fermions and, optionally, Goldstone bosons by multiplications in some coset space. If the field and coordinate transformations are to form the standard extended Poincar\'{e} superalgebra, one can parameterize the coset space in such a way that the transformations of superspace coordinates have standard form, the transformations of the fermionic superfields mimic those of the Volkov-Akulov field \cite{neugold1}, \cite{neugold2} and the odd coordinates of the superspace do not transform with respect to broken supersymmetry. Thus, the first component of the Goldstone fermionic superfield also transforms according to the Volkov-Akulov law, while the bosonic components remain inert with respect to broken supersymmetry. As a result, one can construct a broken supersymmetry invariant action with arbitrary bosonic limit by covariantizing integration measure and derivatives and adding the Wess-Zumino terms and then fix the action completely by invariance with respect to unbroken supersymmetry. One can note that this scheme does not explicitly depend on the number of supersymmetries involved, as it is easy to construct the Volkov-Akulov action for an arbitrary number of Goldstone fermions.

To check feasibility of this idea, it would be desirable to study the simplest possible field theoretic models which realize patterns of supersymmetry breaking different from $1/2$. Thus, we do not concentrate on the already known systems of branes with unusual amounts of preserved supersymmetry but study the three-dimensional models with one unbroken $N=1$ supersymmetry and some $N_0$ number of spontaneously broken supersymmetries, which are realized by the shifts of the Goldstone fermionic fields. Two separate cases can be studied as the Goldstone fermions can belong to either scalar or vector multiplets.

\section{Basic formulae}
Let us begin the study with the system of some $N_0$ scalar multiplets. The superalgebra under consideration is a direct sum of $N_0+1$ copies of $N=1$, $d=3$ superalgebra, with some central charges $Z^i$:
\be\label{N0plus1}
\big\{ Q_\alpha, Q_\beta \big\} = 2\big(\sigma^A \big)_{\alpha\beta} P_A, \;\; \big\{ S^i_\alpha, S^j_\beta \big\} = 2\delta^{ij}\big(\sigma^A \big)_{\alpha\beta} P_A,\;\; \big\{ Q_\alpha, S^i_\beta \big\} = 2\epsilon_{\alpha\beta} Z^i.
\ee
Here, $\alpha,\beta,\ldots=1,2$ and $A,B,\ldots =0,1,2$ are $SO(1,2)$ Majorana spinor and vector indices, respectively, and $i=1,\ldots,N_0$. The $\sigma$-matrices, used here, satisfy the property
\be\label{3dsigmaM2}
\big(  \sigma^A  \big)_{\alpha\gamma}\big( \sigma^B  \big)^{\gamma\beta} = \eta^{AB}\delta_\alpha^\beta + \epsilon^{ABC}\big(\sigma_C \big)_\alpha^\beta, \;\; \epsilon^{ABC} = \epsilon^{[ABC]}, \; \epsilon^{012}=1.
\ee
Spinor and vector indices are raised and lowered by the tensors
\be\label{4dspeps}
\epsilon_{\alpha \beta} = - \epsilon_{\beta \alpha}, \; \epsilon_{\alpha \gamma} \epsilon^{\gamma \beta} =  \delta_{\alpha}^{\beta}, \; \epsilon_{12} = - \epsilon^{12} = 1, \; (\epsilon_{\alpha\beta})^\dagger =\epsilon_{\alpha\beta}, \;\; \eta^{AB} = \mbox{diag}\big( 1,-1,-1  \big).
\ee
The algebra \p{N0plus1} can also be extended by Lorenz $SO(1,2)$ and $SO(N_0)$ generators, which together form the algebra of external automorphisms of \p{N0plus1}.

We consider spontaneous breaking of $N=N_0+1$, $d=3$ supersymmetry to $N=1$, $d=3$, with so far arbitrary $N_0$. For this purpose, it is natural to parameterize the coset element as
\be\label{coset}
g = e^{\im x^A P_A }e^{\theta^\alpha Q_\alpha} e^{\mpsi^{i\alpha}S^i_{\alpha}} e^{\im \mq^i Z^i}.
\ee
Here, $x^A$ and $\theta^\alpha$ are the usual coordinates of the $N=1$, $d=3$ superspace, while $\mq^i$ and $\mpsi^{i\alpha}$ are the $N=1$, $d=3$ superfields that depend on $x^A$, $\theta^\alpha$. The invariant Maurer-Cartan forms can now be easily calculated as
\bea\label{CF}
g^{-1}dg &=& \im \triangle x^A P_A + d\theta^\alpha Q_\alpha + d\mpsi^{i\alpha}S^i_\alpha + \im \triangle \mq^i Z^i, \nn \\
\mbox{where} \;\; \triangle x^A &= & dx^A + \im \big(d\theta^\alpha \, \theta^\beta + d\mpsi^{i\alpha}\,\mpsi^{i\beta}   \big)\big(\sigma^A  \big)_{\alpha\beta}, \;\; \triangle \mq^i = d\mq^i +2\im d\theta^\alpha \, \mpsi^i_\alpha.
\eea
Transformations in the coset space \p{coset} are supposed to be induced by the left multiplication, $g_0 g=g^\prime h$, where $h$ is the element of small subgroup, which contains Lorentz and $SO(N_0)$ generators. In the most interesting cases, unbroken and broken supersymmetries, $h=1$:
\bea\label{cosettrQ}
g_Q &=& e^{\epsilon^\alpha Q_\alpha} \; \Rightarrow \; \delta_Q \theta^\alpha = \epsilon^\alpha,\; \delta_Q x^A = \im \epsilon^\alpha\theta^\beta \big(\sigma^A  \big)_{\alpha\beta}, \; \delta_Q \mpsi^{\alpha i} =0, \; \delta_Q \mq^i =0,\nn \\
\label{cosettrS} g_S &=& e^{\varepsilon^{i\alpha} S^i_\alpha} \; \Rightarrow \; \delta_S \mpsi^{\alpha i} = \varepsilon^{\alpha i },\; \delta_S x^A = \im \varepsilon^{\alpha i}\mpsi^{\beta i} \big(\sigma^A  \big)_{\alpha\beta}, \; \delta_Q \theta^{\alpha } =0, \; \delta_Q \mq^i = \im \varepsilon_\alpha^i \theta^\alpha.
\eea

Using the invariant forms $\triangle x^A$ and $d\theta^\alpha$, one can establish a set of derivatives, which are covariant with respect to $N_0+1$ supersymmetries:
\bea\label{covders}
\nabla_A  =\big( E^{-1}\big)_A{}^B\partial_B, \;\; E_A{}^B = \delta_A^B + \im \partial_A \mpsi^{i\mu}\, \mpsi^{i\nu}\big(  \sigma^B\big)_{\mu\nu}, \nn \\
\nabla_\alpha = D_\alpha -\im \nabla_\alpha \mpsi^{i\mu}\, \mpsi^{i\nu}\big(  \sigma^B\big)_{\mu\nu} \partial_B, \;\; D_\alpha = \frac{\partial }{\partial\theta^\alpha} - \im \theta^\beta \big(\sigma^C  \big)_{\alpha\beta} \partial_C.
\eea
The derivatives \p{covders} satisfy the following commutation relations:
\bea\label{covdercom}
\big\{ \nabla_\alpha, \nabla_\beta \big\} &=& -2\im \big(\sigma^A  \big)_{\alpha\beta} \nabla_A - 2\im \nabla_\alpha \mpsi^{j\mu}\, \nabla_\beta\mpsi^{j\nu} \big(\sigma^C  \big)_{\mu\nu} \nabla_C, \nn \\
\big[ \nabla_A, \nabla_\alpha \big] &=& -2\im \nabla_A \mpsi^{j\mu}\,\nabla_\alpha \mpsi^{j\nu}\big( \sigma^B \big)_{\mu\nu}\nabla_B, \\ \big[ \nabla_A, \nabla_B \big] &=& 2\im  \nabla_A \mpsi^{j\mu} \nabla_B \mpsi^{j\nu} \big( \sigma^B   \big)_{\mu\nu} \nabla_B. \nn
\eea

As the $d\theta^\alpha$ and $\triangle \mq^i$ differential forms are covariant with respect to both supersymmetries, the $d\theta^\alpha$ projection of $\triangle \mq^i$ can be put to zero:
\be\label{IHE}
d\theta^\alpha \big( \nabla_\alpha \mq^i + 2\im \mpsi^i_\alpha  \big) =0 \;\Rightarrow \; \mpsi^i_\alpha = \frac{\im}{2}\nabla_\alpha\mq^i.
\ee
Thus, the fermionic Goldstone fields are not independent and can be expressed in terms of the derivatives of the Goldstone bosons. This is an example of the inverse Higgs effect \cite{invhiggs}.

Studying the $\nabla_\beta$ derivative of relation \p{IHE}, we find that the spinorial derivatives of $\mpsi^{\alpha i}$ are also constrained:
\be\label{IHEcons}
\nabla_\alpha \mpsi_\beta^i + \nabla_\beta \mpsi_\alpha^i = \frac{\im}{2}\big\{ \nabla_\alpha,\nabla_\beta\big\}\mq^i = \big(\sigma^C\big)_{\alpha\beta}\nabla_C \mq^i + \nabla_\alpha\mpsi^{j\mu}\, \nabla_\beta\mpsi^{j\nu} \, \big(\sigma^C\big)_{\mu\nu}\nabla_C \mq^i.
\ee
Substituting here $\nabla_\alpha \mpsi_\beta^i $ as a sum of its antisymmetric and symmetric parts,
\be\label{nablapsidef}
\nabla_\alpha \mpsi_\beta^i = \epsilon_{\alpha\beta}\mA^i + \frac{1}{2}\big( \sigma^C \big)_{\alpha\beta}\mJ_C^i,
\ee
we find that the $\mA^i$ superfields are not constrained by equation \p{IHEcons}, while the vectors $\mJ_A^i$ satisfy the nonlinear equation
\be\label{Jeq}
\mJ_A^i = \Big( 1+ \mA^k \mA^k -\frac{1}{4}\mJ_C^k \mJ^{Ck}   \Big)\nabla_A \mq^i + \frac{1}{2}\mJ_A^k \mJ^{Bk}\nabla_B\mq^i -\epsilon_A{}^{BC}\nabla_B \mq^i \mJ_C^m \mA^m.
\ee

\section{Broken supersymmetry}
The components of $N_0$ scalar $N=1$, $d=3$ multiplets can be found as the first components of each of the mentioned superfields:
\be\label{comps}
q^i = \mq^i|_{\theta\rightarrow 0}, \;\; \psi^{i\alpha} = \mpsi^{i\alpha}|_{\theta\rightarrow 0}, \;\; A^i = \mA^i|_{\theta\rightarrow 0}, \;\; J_A^i = \mJ^i_A|_{\theta\rightarrow 0}.
\ee
Relation \p{Jeq} for the components implies
\be\label{Jeqcomp}
J_A^i = \Big( 1+ A^k A^k -\frac{1}{4}J_C^k J^{Ck}   \Big)\cD_A q^i + \frac{1}{2}J_A^k J^{Bk}\cD_Bq^i -\epsilon_A{}^{BC}\cD_B q^i J_C^m A^m.
\ee
Here, $\cE_A{}^B = E_A{}^B|_{\theta\rightarrow 0}$ and $\cD_A = \big(\cE^{-1}\big)_A{}^B \partial_B$ is the usual broken supersymmetry covariant derivative, which acts on the components. Indeed, as $\theta^\alpha$ do not transform under broken supersymmetry, the active transformation laws $\delta^\star_S f = \delta_S \mf|_{\rightarrow 0}- \delta x^A |_{\rightarrow 0} \partial_A f$ of the components read
\bea\label{compStr}
\delta_S^\star q^i = -U^C \partial_C q^i, \;\; \delta_S^\star \psi^{i\alpha} = \varepsilon^{i\alpha} - U^C \partial_C \psi^{i\alpha}, \;\; \delta_S^\star A^i =  - U^C \partial_C A^i, \nn \\ \delta_S^\star J_A^i =  - U^C \partial_C J_A^i, \;\; U^A =\im  \varepsilon^{k\mu}\psi^{k\nu}\big(\sigma^A\big)_{\mu\nu}.
\eea
It is now easy to check that
\bea\label{EStr}
\delta_S^\star \cE_A{}^B &=& - \partial_A U^C \cE_C{}^B - U^C \partial_C \cE_A{}^B, \;\; \delta_S^\star \big(\cE^{-1}\big)_A{}^B =  \cD_A U^B  - U^C \partial_C \big(\cE^{-1}\big)_A{}^B, \nn \\
 \delta_S^\star \cD_A q^i &=& -U^C \partial_C \cD_A q^i, \;\; \delta_S^\star \det\cE = -\partial_C\big( U^C \,\det\cE \big).
\eea

As one can recognize $A^i$ as auxiliary components of the scalar multiplets, some covariant equations of motion should be provided for them (or, equivalently, some terms in the Lagrangian that contain these fields and are invariant with respect to broken supersymmetry and the shifts of the fields $q^i$). Moreover, the terms in both cases should be dimensionless and Lorentz invariant. The number of terms that satisfy these requirements is very large. To simplify the consideration, let us assume that all the terms in the Lagrangian are also $SO(N_0)$ covariant. Then one can still construct many invariant terms, such as $A^i A^i$, $A^i \cD_B q^i \cD^B q^j A^j $ and terms with higher powers of $\cD_A q^i$, but if no odd powers of $A^i$ are present in the action, they lead to the homogeneous equation of motion of $A^i$, which implies simply $A^i=0$. If one wants to avoid explicit breaking of $SO(N_0)$ invariance, it is not possible to construct a term containing odd powers of $A^i$ for arbitrary $N_0$. Thus, we remove $A^i$ from equation \p{Jeqcomp} by taking $A^i=0$ as an equation of motion. However, the invariant linear term $\sim \epsilon^{ABC}\cD_A q^i \cD_B q^j \cD_C q^k \epsilon^{ijkl}A^l$ can be constructed for $N_0=4$, and this case should be studied separately.

The component approach to the actions with partial spontaneous breaking of supersymmetry involves the construction of the ansatz for the action invariant with respect to broken supersymmetry, and fixing of remaining arbitrariness by demanding its invariance with respect to unbroken supersymmetry. Using the transformations \p{EStr}, it is easy to check that the term
\bea\label{Lmain}
{\cal L}_0 = -\det\cE\, F\big( \tr d , \tr d^2, \tr d^3   \big), \;\; d_{AB} = \cD_A q^i \cD_B q^i, \nn \\ \tr d =d_A{}^A, \; \tr d^2 = d_A{}^B d_B{}^A, \; \tr d^3  = d_A{}^B d_B{}^C d_C{}^A,
\eea
transforms with respect to broken supersymmetry as
\be\label{deltaSLmain}
\delta^\star_S {\cal L}_0 = -\partial_C \big( U^C {\cal L}_0  \big).
\ee
The Wess-Zumino term ${\cal L}_{WZ}$ whose variation with respect to broken supersymmetry contains, in addition to $-\partial_K\big( U^K {\cal L}_{WZ} \big)$, another full derivative, could also be constructed. Indeed, one can check that
\bea\label{LWZ}
{\cal L}_{WZ} &=&\im \det\cE\,\epsilon^{ABC}\cD_A q^i \cD_B q^j \big( \psi^{i\alpha}\cD_C \psi_\alpha^j -  \psi^{j\alpha}\cD_C \psi_\alpha^i \big) \; \Rightarrow \nn \\
\delta^\star_S{\cal L}_{WZ} &=& - \partial_K\big( U^K {\cal L}_{WZ} \big) + \im\det\cE\, \epsilon^{ABC}\cD_A q^i \cD_B q^j \big( \varepsilon^{i\alpha}\cD_C \psi_\alpha^j -  \varepsilon^{j\alpha}\cD_C \psi_\alpha^i \big) =\nn \\
&=& - \partial_K\big( U^K {\cal L}_{WZ} \big) +\im \partial_C \Big(\epsilon^{ABC}\partial_A q^i \partial_B q^j \big( \varepsilon^{i\alpha} \psi_\alpha^j -  \varepsilon^{j\alpha} \psi_\alpha^i \big)\Big).
\eea

\section{Unbroken supersymmetry}
The transformation laws of the first components of the superfields in the case $A^i=0$ can be found with the help of the formula:
\be\label{deltaQ}
\delta_Q^\star f = \epsilon^\gamma D_\gamma \mf |_{\theta\rightarrow 0}= \epsilon^\gamma \nabla_\gamma \mf |_{\theta\rightarrow 0} + H^C \partial_C f, \;\; H^A = \frac{\im}{2} \epsilon^\gamma \psi^i_\gamma J^{Ai} + \frac{\im}{2} \epsilon^{ABC}\epsilon^\mu \psi^{i\nu}J_B^i \big( \sigma_C  \big)_{\mu\nu}.
\ee
Therefore, the transformations of the basic components $q^i$ and $\psi^{i\alpha}$ under unbroken supersymmetry read
\be\label{deltaQqpsi}
\delta_Q^\star q^i = -2\im \epsilon^\gamma \psi_\gamma^i + H^C \partial_C q^i, \;\; \delta_Q^\star \psi^{i\alpha} = \frac{1}{2}\epsilon^\gamma \big( \sigma^C \big)_\gamma{}^\alpha\, J_C^i + H^C \partial_C \psi^{i\alpha}.
\ee
The transformations of the most essential ingredients of the action can now be calculated in a straightforward way:
\bea\label{deltaQEDq}
\delta_Q^\star \cE_A{}^B &=& H^M \partial_M \cE_A{}^B +\partial_A H^M\, \cE_M{}^B - \im \epsilon^\gamma \partial_A \psi^i_\gamma J^{iB} - \im \epsilon^\mu \partial_A \psi^{i\nu}\big( \sigma_D \big)_{\mu\nu}J^i_C\, \epsilon^{BCD}, \nn \\
\delta_Q^\star \det\cE &=& \partial_M \big( H^M  \det\cE \big) -\im \det\cE \epsilon^\gamma \cD_C \psi^i_\gamma J^{iC}- \im \det\cE\epsilon^\mu \cD_B \psi^{i\nu}\big( \sigma_D \big)_{\mu\nu}J^i_C\, \epsilon^{BCD}, \\
\delta_Q^\star \cD_A q^i &=& -2\im \epsilon^\gamma \cD_A \psi_\gamma^i +\im \epsilon^\gamma \cD_A \psi_\gamma^k J^{kB}\cD_B q^i + \im \epsilon^\mu \cD_A \psi^{k\nu}\big( \sigma_D  \big)_{\mu\nu}\, J_C^k\, \cD_B q^i\, \epsilon^{BCD} + H^M \partial_M\cD_A q^i.\nn
\eea

The variation of the action under unbroken supersymmetry transformations is a combination of variations of the main part and of the Wess-Zumino term. Let us study them separately.

As the transformation laws of the $\cD_A q^i$ explicitly contain $J_A^i$ and it is not possible to solve equation \p{Jeqcomp} for $J_A^i$ in a convenient way, let us present the main part of the action as a function of $J_A^i$:
\bea\label{LmainJ}
{\cal L}_0 = -\det\cE\, {\widetilde F}\big( \tr X , \tr X^2 , \tr X^3   \big), \;\; X_{AB} = J_A^i J_B^i, \nn \\ \tr X =X_A{}^A, \; \tr X^2 = X_A{}^B X_B{}^A, \; \tr X^3  = X_A{}^B X_B{}^C X_C{}^A.
\eea
This makes sense as equation \p{Jeqcomp} is linear with respect to $\cD_A q^i$ and can be solved in a straightforward way:
\bea\label{cDqJ}
J_A^i &=& M_A{}^B \cD_B q^i, \;\; M_A{}^B = \Big( 1 - \frac{1}{4} \tr X \Big)\delta_A^B + \frac{1}{2}X_A{}^B \;\; \Rightarrow \cD_A q^i = \big( M^{-1}  \big)_A{}^BJ_B^i,  \\
\cD_A q^i &=&  \frac{1}{\det M}\Big[\Big( 1 + \frac{1}{16} \big(\tr X \big)^2 -\frac{1}{8}\tr X^2    \Big)J_A^i -\frac{1}{2} \Big( 1 + \frac{1}{4}\tr X   \Big)X_A{}^B J_B^i + \frac{1}{4}\big( X^2  \big)_A{}^B J_B^i \Big], \nn \\
\det M &=& 1- \frac{1}{4}\tr X + \frac{1}{16}\big(\tr X \big)^2 + \frac{1}{192}\big(\tr X \big)^3 -\frac{1}{8}\tr X^2\  -\frac{1}{32}\tr X \tr X^2  + \frac{1}{24}\tr X^3 .\nn
\eea
Using this formula, one can express $\tr d$, $\tr d^2 $, $\tr d^3 $ in terms of $\tr X $, $\tr X^2 $, $\tr X^3 $, and also find the variations of $\tr d$, $\tr d^2 $, $\tr d^3 $ in terms of $J_A^i$. Then the transformation laws of $\tr X $, $\tr X^2 $, $\tr X^3 $ under unbroken supersymmetry can be found as a solution of the system of linear equations:
\bea\label{trdtrX}
\delta^\star_Q \tr d = \frac{\partial \tr d}{\partial\tr X  } \delta^\star_Q \tr X  + \frac{\partial \tr d}{\partial\tr X^2  } \delta^\star_Q \tr X^2 + \frac{\partial \tr d}{\partial\tr X^3 } \delta^\star_Q \tr X^3 , \nn \\
\delta^\star_Q \tr d^2 = \frac{\partial \tr d^2}{\partial\tr X  } \delta^\star_Q \tr X  + \frac{\partial \tr d^2 }{\partial\tr X^2  } \delta^\star_Q \tr X^2 + \frac{\partial \tr d^2}{\partial\tr X^3  } \delta^\star_Q \tr X^3 ,  \\
\delta^\star_Q \tr d^3  = \frac{\partial \tr d^3 }{\partial\tr X  } \delta^\star_Q \tr  X  + \frac{\partial \tr d^3 }{\partial\tr X^2  } \delta^\star_Q \tr X^2 + \frac{\partial \tr d^3}{\partial\tr X^3  } \delta^\star_Q \tr X^3.\nn
\eea
The most practical way to perform this calculation is to use a computer analytical calculation package such as {\it Mathematica}. The result, however, is too complicated to be written explicitly. One can still note that 1) terms with $H^A$ are combined into full divergence $\partial_A \big( H^A {\cal L}_0 \big)$ regardless of the explicit form of the function ${\widetilde F}$ and 2) the variation of $\det\cE {\widetilde F}$ can be written as a sum of six linearly independent terms with coefficients that depend on ${\widetilde F}$ and its derivatives:
\bea\label{deltaQLmainterms}
\det\cE\epsilon^\alpha \cD_A \psi_\alpha^i J^{iA}, \;\; \det\cE\epsilon^\alpha \cD_A \psi_\alpha^i X^{AB}J^{i}_B, \;\; \det\cE\epsilon^\alpha \cD_A \psi_\alpha^i  \big(X^2\big)^{AB}J^{i}_B, \nn \\
\det\cE\epsilon^\alpha \cD_A \psi^{i\beta}\big( \sigma_D \big)_{\alpha\beta} J^{i}_C \epsilon^{CDA}, \;\; \det\cE\epsilon^\alpha \cD_B \psi^{i\beta}X^B_A \big( \sigma_D \big)_{\alpha\beta} J^{i}_C \epsilon^{CDA}, \\ \det\cE\epsilon^\alpha \cD_B \psi^{i\beta}\big(X^2\big)^B_A \big( \sigma_D \big)_{\alpha\beta} J^{i}_C \epsilon^{CDA}.\nn
\eea
The terms in the first line can not be combined into full divergence, even in the first approximation in the fermions and when rewritten as functions of $\partial_A q^i$. Therefore, their coefficients should vanish completely. This condition is strong enough to determine the function ${\widetilde F}$ up to a numerical multiplier; we normalize it as
\be\label{tildeF}
{\widetilde F} = -\frac{2 \Big( 1 -\frac{1}{4} \tr X \Big)^2}{\det M}.
\ee
The variation of the main part of the Lagrangian, up to full divergence, can now be written as
\be\label{deltaQLmain}
\delta_Q^\star {\cal L}_0= 2\im \frac{\det\cE }{\det M } \Big( 1 -\frac{1}{4} \tr X  \Big)\, \epsilon^\alpha \cD_A   \psi^{i\beta}M_B{}^A\big( \sigma_D \big)_{\alpha\beta} J^{i}_C \epsilon^{BCD}.
\ee

The variation of the Wess-Zumino term \p{LWZ}
\be\label{LWZ2}
{\cal L}_{WZ} =\im \epsilon^{ABC}\partial_A q^i \partial_B q^j \big( \psi^{i\alpha}\partial_C \psi_\alpha^j -  \psi^{j\alpha}\partial_C \psi_\alpha^i \big)
\ee
with respect to unbroken supersymmetry can be written as
\bea\label{deltaQLWZ}
\delta_Q^\star {\cal L}_{WZ} &=& 2 \im \epsilon^{ABC}\partial_A \delta_Q^\star q^i \, \partial_B q^j \big( \psi^{\alpha i}\partial_C \psi_\alpha^j - \psi^{\alpha j}\partial_C \psi_\alpha^i   \big) + 2 \im \epsilon^{ABC} \partial_A q^i \partial_B q^j \partial_C \big( \psi^{i\alpha}\delta_Q^\star \psi^j_\alpha  \big) +\nn \\&&+4\im \epsilon^{ABC}\partial_A q^i \partial_B q^j \delta_Q^\star\psi^{i\alpha}\, \partial_C \psi^j_\alpha =\nn \\
&=&2\im \epsilon^{ABC} \partial_A \big[ \delta_Q^\star q^i \, \partial_B q^j \big(  \psi^{\alpha i}\partial_C \psi_\alpha^j - \psi^{\alpha j}\partial_C \psi_\alpha^i \big) \big] + 2 \im \epsilon^{ABC} \partial_A q^i \partial_B q^j \partial_C \big( \psi^{i\alpha}\delta_Q^\star \psi^j_\alpha  \big)+ \nn \\
&&+4\im \epsilon^{ABC}\big[ H^M \partial_M q^i \, \partial_A q^j \partial_B \psi^{i\alpha}\partial_C \psi^j_\alpha + \partial_A q^i \, \partial_B q^j \, H^M \partial_M \psi^{i\alpha}\, \partial_C \psi^j_\alpha \big] -\nn \\
&&- 2\im \epsilon^{ABC} \partial_A q^i \, \partial_B q^j \, \epsilon^\alpha \big( \sigma^D  \big)_{\alpha\beta}\partial_C \psi^{j\beta} J_D^i + 8 \epsilon^{ABC} \epsilon^\beta \psi_\beta^i\, \partial_A q^j \, \partial_B \psi^{i\alpha} \partial_C \psi^j_\alpha.
\eea
Looking at this result, one can note that only the last line is essential as terms with $H^M$ can be proven to vanish due to the identity $\epsilon^{ABC}H^M = \epsilon^{MBC}H^A + \epsilon^{AMC}H^B +\epsilon^{ABM}H^C$, and the other two terms are just total divergences.

The first term in the last line of \p{deltaQLWZ} can be transformed further. Restoring $\det\cE$ and the covariant derivatives in it and taking into account that $\cD_A q^i = \big( M^{-1}\big)_A{}^B\, J_B^i$, one can find that
\bea\label{deltaQLWZ2}
\im \det\cE \epsilon^{ABC} \cD_A q^i \cD_B q^j \epsilon^\alpha \big( \sigma^D \big)_{\alpha\beta}\cD_C \psi^{j\beta} J_D^i = \im \frac{\det\cE}{\det M} \epsilon^{ABC}J_A^i J_D^i J_B^j \epsilon^\alpha \big( \sigma^D \big)_{\alpha\beta}M_C{}^K \cD_K \psi^{j\beta} = \nn \\
= 2\im \det\cE \epsilon^{ABC} \cD_B q^j \epsilon^\alpha \big( \sigma_A \big)_{\alpha\beta}\cD_C \psi^{j\beta} +2\im \frac{\det\cE}{\det M} \Big( 1 -\frac{1}{4}\tr X   \Big)\epsilon^{ABC} \epsilon^{\alpha}J_A^i \big( \sigma_B \big)_{\alpha\beta} M_C{}^K \cD_K \psi^{j\beta}.
\eea
Here, we used the fact that $J_A^i\, J_D^i = X_{AD} = 2 M_{AD} -2\Big( 1 -\frac{1}{4}\tr X   \Big)\eta_{AD} $. Moreover, the term with the derivative of $q^i$ is combined with the last term in the last line of \p{deltaQLWZ} into full divergence:
\bea\label{deltaQLWZ3}
-4\im \epsilon^{ABC} \partial_B q^j \epsilon^\alpha \big( \sigma_K \big)_{\alpha\beta} \cE_A{}^K \partial_C \psi^{j\beta} + 8 \epsilon^{ABC} \epsilon^\beta \psi_\beta^i\, \partial_A q^j \, \partial_B \psi^{i\alpha} \partial_C \psi^j_\alpha = \nn \\
=-4\im  \epsilon^{ABC}\big( \sigma_A \big)_{\alpha\beta}\epsilon^\alpha \partial_C \psi^{j\beta} \, \partial_B q^j -2 \epsilon^{ABC} \partial_B q^j \epsilon^\alpha \partial_A \big( \psi^i_\lambda \psi^{i\lambda} \big) \partial_C \psi^{j}_\alpha.
\eea

Therefore, the variation of the Wess-Zumino term \p{deltaQLWZ}, up to the total derivative, could be written just as
\be\label{deltaQLWZ4}
\delta_Q^\star {\cal L}_{WZ} = -4\im \frac{\det\cE}{\det M} \Big( 1 -\frac{1}{4}\tr X  \Big)\epsilon^{ABC} \epsilon^{\alpha}J_A^i \big( \sigma_B \big)_{\alpha\beta} M_C{}^K \cD_K \psi^{i\beta}.
\ee
Therefore, the combination ${\cal L}_0 + 1/2\,{\cal L}_{WZ}$ is invariant with respect to unbroken supersymmetry.

\section{Comments on the general action}
The action of $N_0$ scalar $N=1$, $d=3$ supermultiplets, which is invariant with respect to one unbroken and $N_0$ spontaneously broken supersymmetries as well as the $SO(N_0)$ group, reads
\be\label{genN0action}
S =- \int d^3 x \det\cE \frac{2 \Big( 1 -\frac{1}{4} \tr X  \Big)^2}{\det M} +\frac{\im}{2}\int d^3 x \,\det\cE\,\epsilon^{ABC}\cD_A q^i \cD_B q^j \big( \psi^{i\alpha}\cD_C \psi_\alpha^j -  \psi^{j\alpha}\cD_C \psi_\alpha^i \big).
\ee
Its structure is uniquely fixed by the mentioned requirements.

It is worth noting that \p{genN0action} actually depends on $q$, not on $J$, while rewriting it in terms of $\cD_A q^i$ is not easy in general. This can be done most easily in the particular cases of $N_0=1$ and $N_0=2$, when relation \p{Jeqcomp} can be solved for $J_A^i$ analytically. For $N_0=1$, we find the action of the membrane in $d=4$:
\bea\label{N01action}
J_A =\big( 1-\frac{1}{4}J_B J^B \big)\cD_A q +\frac{1}{2}J_A J^B \cD_B q \;\Rightarrow \; \cD_A q =\frac{J_A}{1+\frac{1}{4}J_B J^B}, \; J_A = \frac{2\cD_A q}{1+ \sqrt{1-\cD_B q \cD^B q}}, \nn \\
S_{N_0=1} =-\int d^3 x \det\cE \Big( 1 + \sqrt{1-\cD_A q\cD^A q} \Big).
\eea
These formulae can be obtained by setting $\tr X^2 = \big(\tr X\big)^2$, $\tr X^3 = \big(\tr X\big)^3$ in the general formulae above. This result, obviously, coincides with one obtained in \cite{cosetact}.

For $N_0=2$, one still can express $\tr X^3 $ in terms of the traces of the lower powers of $X$ by noting that $\det X =0$. Therefore, introducing the notation
\be\label{N02qpsi}
q^1 = q + \bq, \; q^2=\im(q-\bq), \; \psi^1_\alpha = \frac{1}{\sqrt{2}}\big( \psi_\alpha+\bpsi_\alpha \big), \; \psi^2_\alpha = \frac{\im}{\sqrt{2}}\big( \psi_\alpha-\bpsi_\alpha \big),
\ee
we obtain exactly the component action of the membrane in $D=5$ \cite{actD5}:
\bea\label{N02trX3}
S_{N_0=2} &=& -2\int d^3 x\, \det\cE \frac{1-\frac{1}{4}\tr X}{1+\frac{1}{16}\big(\tr X\big)^2 - \frac{1}{8}\tr X^2} + \nn\\&&+\frac{\im}{2}\int d^3 x \,\det\cE\,\epsilon^{ABC}\cD_A q^i \cD_B q^j \big( \psi^{i\alpha}\cD_C \psi_\alpha^j -  \psi^{j\alpha}\cD_C \psi_\alpha^i \big)=\nn \\
&= &-\int d^3 x\, \det\cE \big( 1+\sqrt{\det{\big(\eta_{AB}-2\cD_A q\cD_B \bq -2\cD_A \bq \cD_B q \big)}}  \big) - \nn \\&&  -2\im\int d^3 x \,\det\cE\,\epsilon^{ABC}\cD_A q \cD_B \bq \big( \psi^{\alpha}\cD_C \bpsi_\alpha -  \bpsi^{\alpha}\cD_C \psi_\alpha \big).
\eea
Thus, this action is invariant with respect to additional hidden unbroken $N=1$, $d=3$ supersymmetry and we again deal with breaking of just one half of the supersymmetry.

For $N_0\geq3$, the effects of the unusual fraction of supersymmetry breaking become nontrivial and the action \p{genN0action} can not be written in standard Nambu-Goto form, $\sqrt{\det{\big(\eta_{AB}-\cD_A q^i \cD_B q^i \big)}}$ is not even a rational fraction in $J$. While it is not possible to solve equation \p{Jeqcomp} for $J_A^i\big( \cD_B q^j \big)$ for general $N_0$ in closed form, one still can reduce the problem to one scalar equation and obtain the general matrix structure. Let us slightly rewrite equation \p{Jeqcomp} as
\be\label{Jeqcomp2}
J_A^i = \Lambda\cD_A q^i + \frac{1}{2}J_A^k J^{Bk}\cD_Bq^i, \;\; \Lambda = 1-\frac{1}{4}J_C^k J^{Ck}
\ee
and use $\Lambda$ as a parameter. Now substituting $J_A^i = M_A{}^B\big( d \big) \cD_B q^i$,
\be\label{Jeqcomp3}
\Big( \frac{1}{2}M_A{}^B d_{BC} M^{CD} - M_A{}^D +\Lambda \delta_A{}^D   \Big)\cD_D q^i =0.
\ee
If $N_0\geq 3$, this equation implies that the matrix in brackets is zero. Also noting that $M$ and $d$ commute, one can solve this simple quadratic equation. The solution, which is not singular when $\cD_B q^i \rightarrow 0$, is
\be\label{Mdqsol}
M_A{}^B = 2\Lambda \Big( \frac{1}{1+\sqrt{1-2\Lambda d}}   \Big)_A{}^B.
\ee
As $\tr M =3-\frac{1}{4}J_A^i J^{Ai} =2+\Lambda$, the parameter $\Lambda$ can be found as a function of $\cD_A q^i$ by solving the equation
\be\label{Lambdaeq}
2+\Lambda = 2\Lambda \tr\Big( \frac{1}{1+\sqrt{\det(1-2\Lambda d})}   \Big).
\ee
Finally, one can explicitly write the bosonic part of the general Lagrangian \p{genN0action} as
\be\label{LgendqLambda}
\cL =-\frac{1}{4\Lambda}\det\big( 1+\sqrt{1-2\Lambda d}  \big).
\ee

Unfortunately, it is not known how to solve equation \p{Lambdaeq} nonperturbatively. The function $\Lambda$ and the bosonic Lagrangian still can be written as power series
\bea\label{Lgendq}
\Lambda &=& 1+ \Big[ -\frac{1}{4}\tr d  \Big] + \Big[ -\frac{1}{4}\tr d^2 + \frac{1}{8}\big(\tr d\big)^2  \Big] + \Big[ -\frac{5}{16}\tr d^3 +\frac{5}{16}\tr d^2\, \tr d   -\frac{5}{64}\big(\tr d\big)^3  \Big]+ \nn \\&&+
\Big[ -\frac{1}{32} \big(\tr d^2\big)^2 -\frac{11}{96} \tr d \, \tr d^3 + \frac{7}{64}\tr d^2 \big( \tr d\big)^2 -\frac{7}{384} \big(\tr d\big)^4  \Big]+ \ldots,  \\
\cL &=& -2+\Big[ \frac{1}{2}\tr d  \Big] + \Big[ \frac{1}{4} \tr d^2   - \frac{1}{8} \big(\tr d\big)^2 \Big]
+ \Big[ \frac{1}{24} \big(\tr d  \big)^3-\frac{3}{16} \tr d  \, \tr d^2 +\frac{5}{24} \tr d^3 \Big] +\nn \\ && +\Big[ \frac{7}{384} \big(\tr d  \big)^4 -\frac{3}{32} \big(\tr d  \big)^2\tr d^2  +\frac{1}{32}\big(  \tr d^2   \big)^2 +\frac{1}{12}\tr d \,\tr d^3  \Big]+\ldots\nn
\eea

\section{Special case $N_0=4$}
As mentioned before, the case $N_0=4$ offers a specific possibility of combining $SO(N_0)$ invariance with the nontrivial equation of motion for the auxiliary field. We assume that this equation can be cast into the form
\be\label{N04Ai}
A^i = \Phi \epsilon^{ABC}\epsilon^{ijkl}J_A^j J_B^k J_C^l.
\ee
Let us note that up to the function $\Phi$ this relation is the most general possible, as $J_B^i A^i =0$ for \p{N04Ai} and it can not be multiplied by matrices like $J^{Am}J_A^i$. Indeed, such a multiplication leads to the expression
\be\label{antisym4}
J_A^i J_B^j J_C^k J_D^l \epsilon^{ijkl}=0,
\ee
being antisymmetric in $A,B,C,D$. The property $J_B^i A^i =0$ also simplifies the relation between $J_A^i$ and $\cD_B q^i$:
\be\label{N04Jeqcomp}
J_A^i ={\widetilde M}_{A}{}^B \cD_B q^i, \;\;  {\widetilde M}_{A}{}^B=\Big( 1 -\frac{1}{4}\tr X +36 \Phi^2 \det X  \Big)\delta_A^B + \frac{1}{2}X_A{}^B,
\ee
as the $\epsilon^{ABC}$-term is now absent.

With the $\epsilon^{ijkl}$ symbol at hand, one can construct one more Wess-Zumino term in addition to the general one \p{LWZ}:
\be\label{N04LWZ}
\cL_{WZ2}=\im \det\cE\,\epsilon^{ABC}\epsilon^{ijkl}\cD_A q^i \cD_B q^j \psi^{k\alpha}\cD_C \psi_\alpha^l.
\ee
The proof of its invariance with respect to broken supersymmetry is completely analogous to the proof of invariance of the general term \p{LWZ}. The ansatz for the Lagrangian in the case $N_0=4$, therefore, reads
\be
{\cal L}_{N_0=4} = -\det\cE F(J) +2\im   C_1\det\cE\,\epsilon^{ABC}\cD_A q^i \cD_B q^j  \psi^{i\alpha}\cD_C \psi_\alpha^j + \im C_2 \det\cE\,\epsilon^{ABC}\epsilon^{ijkl}\cD_A q^i \cD_B q^j \psi^{k\alpha}\cD_C \psi_\alpha^l.
\ee

Unbroken supersymmetry transformations can be treated using the same idea as in the previous section. The transformation laws of $q^i$ and $\psi^{i}_\alpha$ under unbroken supersymmetry now read
\bea\label{N04Qtr}
\delta_Q^\star q^i = -2\im \epsilon^\gamma \psi_\gamma^i + {\widetilde H}^C \partial_C q^i, \;\; \delta_Q^\star \psi^{i\alpha} = \frac{1}{2}\epsilon^\gamma \big( \sigma^C \big)_\gamma{}^\alpha\, J_C^i -\epsilon^\alpha \Phi \epsilon^{ijkl}\epsilon^{ABC}J_A^j J_B^k J_C^l+ {\widetilde H}^C \partial_C \psi^{i\alpha}, \nn \\
{\widetilde H}^A = -\im \Phi \epsilon^\mu \psi^{i\nu}  \epsilon^{ijkl}\epsilon^{PQR}J_P^j J_Q^k J_R^l \big(\sigma^A \big)_{\mu\nu} + \frac{\im}{2} \epsilon^\gamma \psi^i_\gamma J^{Ai} + \frac{\im}{2} \epsilon^{ABC}\epsilon^\mu \psi^{i\nu}J_B^i \big( \sigma_C  \big)_{\mu\nu}.
\eea
New linearly independent terms appear in the variation of the action, in addition to \p{deltaQLmainterms}:
\bea\label{N04newterms}
\epsilon^{ABC}\epsilon^{ijkl}J_A^j J_B^k J_C^l\epsilon^\mu \big( \sigma^K  \big)_{\mu\nu}\cD_K \psi^{k\nu}, \;\; \epsilon^{ABC}\epsilon^{ijkl}J_A^j J_B^k J_C^l\epsilon^\mu \big( \sigma^K  \big)_{\mu\nu}\cD_L \psi^{k\nu} X_K^L,\nn \\
\epsilon^{ABC}\epsilon^{ijkl}J_A^j J_B^k J_C^l\epsilon^\mu \big( \sigma^K  \big)_{\mu\nu}\cD_L \psi^{k\nu} \nn \big(X^2\big)_K^L.
\eea
Subsequent analysis in the lowest approximation in fermions shows that to make the system invariant with respect to unbroken supersymmetry, the following two conditions should be satisfied:
\bea\label{N04FPhi}
F =\frac{1}{\det {\widetilde M}}\Big( 4 C_1 \big(1-\frac{1}{4}\tr X    \big)^2 +18 \det X \Phi \big(C_2 +8C_1  \big)\big(1-\frac{1}{4}\tr X \big)   \Big), \\
-C_2 + C_1 \Phi\big( 24   - 6 \tr X  + 864 \det X \Phi^2 \big)=0.\nn
\eea
In contrast to the usual case, the Lagrangian is not a rational function of the $J_A^i$ variables and the constants that accompany the Wess-Zumino terms are not fixed (one of them is irrelevant as it would be fixed by properly normalizing the action). With the special choice of the constants, one can find that the bosonic limit of Lagrangian is the one of the membrane in $D=7$:
\be\label{D7membr}
C_1 =\frac{1}{2}, \; C_2 = \pm\frac{1}{2} \; \Rightarrow \; \cL \sim - 1 -\sqrt{\det\big( \delta_A^B -\partial_A q^i \partial^B q^i  \big)}.
\ee
In this case, one can expect enhancement of unbroken supersymmetry.

\section{Vector multiplets?}
Obviously, one can dualize the scalar fields in the action \p{genN0action} and obtain the action of $N_0$ electromagnetic fields. However, to properly check that the resulting system possesses broken and unbroken supersymmetries, it is desirable to construct it from the first principles. This, however, appears to be problematic.

The standard $N=1$, $d=3$ vector multiplet in the case of one spontaneously broken supersymmetry is described by a spinor superfield $\mpsi^\alpha$ subjected to the irreducibility condition $\nabla_\alpha\mpsi^\alpha=0$ \cite{D2brane}. It is natural to generalize this condition to the case of $N_0$ broken supersymmetries as $\nabla_\alpha\mpsi^{\alpha i}=0$.

It is worth noting that if one wishes to keep $N_0$ arbitrary and the whole system $SO(N_0)$ invariant, it is not possible to modify this relation. Indeed, one can try to introduce some function of $\nabla_\alpha\mpsi^{\beta j}$ to the right-hand side. Let us note that this function should depend on the traceless part of  $\nabla_\alpha\mpsi^{\beta j}$, as the terms with trace parts are combined into the term $N^{ij}\nabla_\alpha\mpsi^{\alpha j}$ and the matrix $N^{ij}$ can be factorized out. As terms with structure $N^{ij}\nabla_\alpha\mpsi^{\alpha j}$ should be avoided, the only candidate cubic term $\nabla_\alpha\mpsi^{\beta k} \nabla_\beta\mpsi^{\gamma k} \nabla_\gamma \mpsi^{\alpha i} $ is zero as $\nabla_\alpha\mpsi^{\beta k} \nabla_\beta\mpsi^{\gamma k}\sim \delta_\alpha^\gamma$ if $\nabla_\alpha\mpsi^{\beta k}$, $\nabla_\beta\mpsi^{\gamma k}$ are traceless. The terms with higher powers of $\nabla_\alpha\mpsi^{\beta k}$ also reduce to zero for the same reasons.

It turns out, however, that the introduced condition is not consistent. It should be used to find the proper Bianchi identity for the field strength. In the bosonic limit it leads to the following condition on the bosonic component of the multiplet:
\be\label{nabla2nablapsi}
\nabla^\beta \nabla_\beta \nabla_\alpha\mpsi^{\alpha i} =0 \; \Rightarrow \; \big\{ \nabla_\alpha, \nabla_\beta\big\} \nabla^\alpha \mpsi^{i\beta}+\mbox{fermions} =0.
\ee
In the vector notation the first component of \p{nabla2nablapsi} reads
\be\label{BI}
\partial_A V^{Ai}\big( 1 - \frac{1}{4}V_B^j V^{Bj}\big) + \frac{1}{2}V^{Aj} V^{Bj}\partial_A V_{B}^i=0, \;\; \nabla_\alpha \mpsi_\beta^i |_{\theta\rightarrow 0} = \frac{1}{2}\big(\sigma^A\big)_{\alpha\beta}V^i_A.
\ee
Contrary to the expectation, this identity is not equivalent to the usual one $\partial_A F^{A i}=0$ and the inconsistency appears already in the cubic approximation in $V$. Indeed, the last term in identity \p{BI} can be rewritten as
\be\label{BIlastterm}
\partial_A\big( V^{Aj} V^{Bj} V_{B}^i \big) - \partial_A V^{Aj}\,  V^{Bj} V_{B}^i - V^{Aj }V^{Bi} \partial_A V_B^j.
\ee
The first term in \p{BIlastterm} contributes to the field strength and the second one can be compensated by multiplying the identity by a suitable matrix with indices $i,j$. The last one, however, could not be compensated unless the equation of motion for the field strength $\epsilon^{ABC}\partial_B V_C^i+\ldots =0$ is taken into account.

The natural way to solve this problem is to assume that the system under consideration can be defined only on-shell, refuse to use the superfield formalism completely and try to formulate this system in terms of fermionic fields and vector potentials. However, even this approach, with some moderate assumptions, seems to be unable to produce the correct transformation laws.

To find the transformation laws of fermions and vector potentials, a good starting point would be to consider broken supersymmetry. It would be natural to keep the broken supersymmetry transformation law of the fermions, obtained within the coset approach, as it is valid for any number of broken supersymmetries
\be\label{dSpsi}
\delta^\star_S \psi_\alpha^i = \varepsilon_\alpha^i - U^C\partial_C \psi_\alpha^i,  \;\; U^A = \im \varepsilon^{j\mu}\psi^{j\nu}\big( \sigma^A \big)_{\mu\nu}.
\ee
As for the vector potential, it is natural to assume that it does not experience an inhomogeneous shift under broken supersymmetry and its transformation law combines a standard term that is induced by the variation of the coordinates with a compensating term to make the whole variation gauge invariant:
\be\label{dSA}
\delta^\star_S {\cA}_A^i=U^B\partial_A\cA_B^i - U^B \partial_B \cA_A^i =  \epsilon_{ABC} U^B {\cal F}^{C i}, \;\; {\cal F}^{Ai}=\epsilon^{ABC}\partial_B {\cA}_C^i.
\ee
The algebra of these transformations closes on both fermions and potentials without any use of the equations of motion:
\be\label{comdSpsiA}
\big[\delta^\star_{S1}, \delta^\star_{S2}\big]\psi_\alpha^i =-2\im \varepsilon^{\mu k}_2 \varepsilon^{\nu k}_1 \big( \sigma^A \big)_{\mu\nu} \partial_A \psi^i_\alpha \; \mbox{and} \; \big[\delta^\star_{S1}, \delta^\star_{S2}\big]{\cA}_A^i = 2\im \varepsilon^{\mu k}_2 \varepsilon^{\nu k}_1 \big( \sigma^B \big)_{\mu\nu} \big( \partial_A {\cA}_B^i - \partial_B {\cA}_A^i \big).\nn
\ee
The commutator of broken supersymmetry transformations, acting on $A^i$, produces not only a shift but also a gauge transformation. As the transformation laws of fermions are standard, the broken supersymmetry covariant derivative should be defined as $\cD_A = \big( \cE^{-1}  \big)_A{}^B \partial_B$. Also note that the quantity $F^{Ai}$ is invariant with respect to broken supersymmetry
\be\label{invFS}
F^{Ai} = \frac{1}{\det\cE} \cE_B{}^A \cF^{Bi}, \;\; \delta^\star_S F^{Ai} = -U^C\partial_C F^{Ai}.
\ee

The next natural point would be to find the general ansatz for unbroken supersymmetry variations, which would commute with broken supersymmetry. The structure of the fermionic field variation  can be fixed easily by assuming that $\big[ \delta^\star_S, \delta_Q^\star  \big]\psi^i_\alpha =0$. At first, one can find $\delta^\star_S \delta_Q^\star \psi^i_\alpha$
\be\label{comdSdQpsi}
\delta^\star_S \delta_Q^\star \psi^i_\alpha = \delta_Q^\star\delta^\star_S  \psi^i_\alpha = -\im \varepsilon^{j\mu} \delta_Q^\star \psi^{j\nu} \big( \sigma^B  \big)_{\mu\nu}\partial_B \psi^i_\alpha - U^A \partial_A \delta_Q^\star\psi^i_\alpha.
\ee
Now it is straightforward to show that
\be\label{hatdpsi}
\delta^\star_S \widehat{ \delta_Q \psi^i_\alpha} = -U^C \partial_C \widehat{ \delta_Q \psi^i_\alpha}, \;\; \widehat{ \delta_Q\psi^i_\alpha} = \delta_Q^\star \psi^i_\alpha + \im \psi^{j\mu} \delta_Q^\star \psi^{j\nu} \big( \sigma^B  \big)_{\mu\nu}\cD_B \psi^i_\alpha.
\ee
The quantity $\widehat{ \delta_Q \psi^i_\alpha}$ is a broken supersymmetry invariant and should be some function of the unbroken supersymmetry parameter $\epsilon^\alpha$ and the covariantized field strength $F^{Ai}$. The complete variation $\delta_Q^\star \psi^i_\alpha$ can be expressed in terms of it,
\be\label{dQpsi}
\delta_Q^\star \psi^i_\alpha = \widehat{ \delta_Q \psi^i_\alpha} -H^C \partial_C \psi^i_\alpha, \;\; H^A = \im \psi^{j\mu}\widehat{ \delta_Q \psi^{j\nu}} \big( \sigma^A \big)_{\mu\nu}.
\ee
It again has the usual structure, and the relation $\big[ \delta^\star_S,\delta_Q^\star  \big]\psi_\alpha^i=0$ is satisfied without any reference to the equations of motion.

Now its natural to assume that the transformation law of the vector potential can be written as
\be\label{dQAgen}
\delta^\star_Q \cA_A^i = \widehat{ \delta_Q \cA_A^i } + \epsilon_{ABC}H^B \cF^{Ci}.
\ee
Commuting it with the broken supersymmetry law, one can obtain that
\be\label{comdSdQA}
\big[ \delta_Q^\star,\delta^\star_S  \big]\cA_A^i = -\delta^\star_S \big(\widehat{ \delta_Q  \cA_A^i}   \big)+ U^B \partial_A  \big(\widehat{ \delta_Q  \cA_B^i}   \big) - U^B \partial_B \big(\widehat{ \delta_Q  \cA_A^i}  \big)+ \partial_A \big( U^B H^C \cF^{Di}\epsilon_{BCD}  \big).
\ee
To close the algebra properly, the right hand side should reduce to a gauge transformation, and the last term already has this form. To reconstruct $\widehat{ \delta_Q \cA_A^i}$ completely, one should note that the leading term in it should coincide with the transformation law of $\cA$ in the free case, $\widehat{ \delta_Q^\star  \cA_A^i}\approx \im \epsilon^\alpha \psi^{\beta i}\big( \sigma_A \big)_{\alpha\beta}$. Thus, relation \p{comdSdQA} should be used to restore higher order fermionic terms in the transformation law.

The ansatz for the transformation law with the most general cubic term that does not assume the equations of motion could be, up to a gauge transformation, written as
\be\label{deltaQAans}
\widehat{{\delta}_Q {\cal A}_A^i} = \im \epsilon^\alpha\psi^{\beta i}\big( \sigma_A \big)_{\alpha\beta} + a \, \epsilon^\alpha \partial_A \psi^i_\alpha \, \psi^{j\beta}\psi_\beta^j + b\, \epsilon^\alpha \psi^{i\beta} \psi^j_\alpha \partial_A \psi_\beta^j.
\ee
Indeed, for any combination of $\epsilon\psi^i\psi^j\psi^j$ and the derivative with the external index $A$ one can employ a gauge transformation to make sure that the derivative acts on the fermion with index $j$. Then only three following combinations could be devised:
\be\label{deltaQAans2}
\epsilon^\alpha \psi^{i}_\alpha \psi^{j\beta}\partial_A \psi^j_\beta, \;\; \epsilon^\alpha \psi^{i\beta} \psi^{j}_\alpha \partial_A \psi^j_\beta, \;\; \epsilon^\alpha \psi^{i\beta} \psi^{j}_\beta \partial_A \psi^j_\alpha.
\ee
The third one is algebraically related to two others and can be excluded. Then the first can be integrated by parts to obtain \p{deltaQAans}.

Substituting it into \p{comdSdQA}, one can find that the following equation should be satisfied:
\bea\label{comdSdQAeq}
-\im \epsilon^\alpha\varepsilon^{\beta i}\big( \sigma_A \big)_{\alpha\beta}+\epsilon^\alpha \varepsilon^{j}_{\alpha}\psi^{\beta j}\partial_A \psi_\beta^i + \epsilon^\alpha \varepsilon^{j\beta} \psi^j_\alpha \partial_A \psi_\beta^i  - \nn \\ -2 a \epsilon^\alpha  \varepsilon^{j\beta}\psi_\beta^j \partial_A \psi^i_\alpha - b \epsilon^\alpha \varepsilon^{i\beta} \psi_\alpha^j \partial_A \psi_\beta^j +b \epsilon^\alpha\varepsilon_\alpha^j \psi^{i\beta}\partial_A \psi^j_\beta = \partial_A f^i.
\eea
The constant term can be presented as a derivative of $-\im \epsilon^\alpha\varepsilon^{\beta i}\big( \sigma_A \big)_{\alpha\beta}x^A$. The term $b \epsilon^\alpha \varepsilon^{i\beta} \psi_\alpha^j \partial_A \psi_\beta^j$ is not a derivative and has no analogues, and, therefore, $b=0$. Finally, representing
\be\label{epsepspsipsi}
\epsilon^\alpha \varepsilon^{\beta j} \psi^j_\alpha \partial_A \psi^i_\beta = \epsilon^\alpha \varepsilon^{ j}_\alpha \psi^{j\beta} \partial_A \psi^i_\beta + \epsilon^\alpha \varepsilon^{\beta j} \psi^j_\beta \partial_A \psi^i_\alpha,
\ee
one can find that the result is not a full derivative regardless of $a$. Thus, the commutator of $Q$ and $S$ transformations, acting on $\cA_A^i$, can not be closed properly. Even the use of the equations of motion could not help here. The equation of motion of the fermionic field in the lowest approximation implies that $\partial_A\psi^i_\alpha \big( \sigma^A  \big)_{\beta\gamma} $ is symmetric in $\alpha$, $\beta$, $\gamma$. As the terms that should be compensated in \p{comdSdQAeq} have a derivative with index $A$, we should consider combinations which contain $\partial_{\alpha\mu}\psi^j_\beta \big( \sigma_A \big)^{\alpha\beta}$. Four of them can be found,
\bea\label{epsepspsipsi2}
\epsilon^\alpha \psi^{i\beta}\psi^j_\beta\partial_{\alpha\mu}\psi^j_\nu \big( \sigma_A \big)^{\mu\nu}, \; \; \epsilon^\alpha \psi^{i}_\alpha \psi^{j\beta}\partial_{\beta\mu}\psi^j_\nu \big( \sigma_A \big)^{\mu\nu}, \;\;
\epsilon^\alpha \psi^{i\beta}\psi^j_\alpha\partial_{\beta\mu}\psi^j_\nu \big( \sigma_A \big)^{\mu\nu}, \; \; \epsilon^\alpha \psi^{j\beta}\psi^{j}_{\beta}\partial_{\alpha\mu}\psi^i_\nu \big( \sigma_A \big)^{\mu\nu}.
\eea
Their variations, after applying the equation of motion, read
\bea\label{epsepspsipsi3}
2 \epsilon^\alpha \varepsilon^{i\beta}\psi^j_\beta \partial_A \psi^j_\alpha + 2\epsilon^\alpha \varepsilon^{j\beta}\psi^i_\beta \partial_A \psi^j_\alpha, \;\;
2 \epsilon^\alpha \varepsilon^{i}_\alpha \psi^{j\beta} \partial_A \psi^j_\beta -2 \epsilon^\alpha \varepsilon^{j\beta}\psi^i_\alpha \partial_A \psi^j_\beta, \nn \\
2 \epsilon^\alpha \varepsilon^{i\beta}\psi^j_\alpha \partial_A \psi^j_\beta -2 \epsilon^\alpha \varepsilon^{j}_\alpha \psi^{i\beta} \partial_A \psi^j_\beta, \;\;
4 \epsilon^\alpha \varepsilon^{j\beta}\psi^j_\beta \partial_A \psi^i_\alpha.
\eea
Regardless of arbitrary coefficients taken in front of these combinations, one can not obtain a full derivative by adding them to \p{comdSdQAeq}.

\section{Conclusion}

The field theory side of non-half breaking of global supersymmetry remains a relatively unexplored subject. To test whether it is possible to construct a usual field theory action with various patterns of supersymmetry breaking, we have considered the systems of multiple scalar and vector $N=1$, $d=3$ Goldstone superfields.

In a simpler case of $N_0$ scalar superfields, it was determined with the usual method of nonlinear realizations that $N_0$ fermions enter into the action as the standard Volkov-Akulov matrix $\cE_A{}^B = \delta_A{}^B -\im \psi^{i\alpha}\partial_A \psi^{i\beta}\big( \sigma^B\big)_{\alpha\beta}$, as well as the Wess-Zumino term. By assuming overall $SO(N_0)$ invariance of the action, which in the case of arbitrary $N_0$ uniquely selects $\nabla^\alpha \nabla_{\alpha} \mq^i|_{\theta\rightarrow 0}=0$ as the equation of motion of the auxiliary field, the unbroken supersymmetry transformation laws were found. They determine the action uniquely. In the case of $N_0=1$ and $N_0=2$, they appear to be the usual static gauge Nambu-Goto actions of the membranes in $D=4$ and $D=5$, respectively. For $N_0 \geq 3$ the action is not of Nambu-Goto type, though its general structure remains universal. Most remarkably, it exists for arbitrary $N_0$. Some peculiarity appears in the case of $N_0=4$ when it is possible to consider the $SO(N_0)$ invariant action with the nontrivial equation of motion of the auxiliary field. This relaxes the requirements on the action and allows one, with proper choice of the parameters, to obtain the action of the membrane in $D=7$ (in the bosonic limit).

Contrary to the result above, it appears to be impossible to construct an action of an arbitrary number of vector multiplets, at least $SO(N_0)$ invariant one. In this case, one should derive the Bianchi identity for the field strength as a consequence of the irreducibility conditions, but the only possible condition implies the identity that could not be brought to the standard form without the use of the equations of motion. Moreover, further analysis of this system, formulated in terms of the vector potential, shows that it is not possible to find the transformation law of the potential which would commute with broken supersymmetry.

As further development of this topic, it would be of interest to construct higher dimensional analogs of scalar systems, which seems to be a rather plausible possibility. Also, it would be desirable to relax the requirement of $SO(N_0)$ symmetry. When it comes to the vector multiplets, additional possibilities are to consider the actions for some particular values of $N_0$, as well as hybrid systems that contain scalars and one or two vector multiplets. Next task could be adaptation of this framework to describe particular systems with unusual patterns of supersymmetry breaking, which are abundant in the literature. Also, it would be interesting to construct a superfield action using the ideas of paper \cite{Ngold}. As a last point, it would be interesting to know whether models for arbitrary $N_0$ admit a string theory interpretation.

\section*{Acknowledgements}
This work was supported by the RFBR, grant 18-52-05002 Arm\_a.



\end{document}